\documentclass[epj]{svjour}
\usepackage{amssymb}
\usepackage{url}
\usepackage{graphics}
\usepackage[normalem]{ulem}   
\usepackage{color}  
\begin{document}
\title{Energy transfer from intense laser pulse to dielectrics in time-dependent density functional theory}

\author{Atsushi Yamada \and Kazuhiro Yabana
}                     
%
%
\institute{Center for Computational Sciences, University of Tsukuba, 1-1-1 Tennodai, Tsukuba, Ibaraki 305-8577 Japan}
%
\date{Received: date / Revised version: date}
%
\abstract{
Energy transfer processes from a high-intensity ultrashort laser pulse to electrons in simple dielectrics, silicon, diamond, and $\alpha$-quartz
are theoretically investigated by first-principles calculations based on time-dependent density functional theory (TDDFT).
Dependences on frequency as well as intensity of the laser pulse are examined in detail, making a comparison with the Keldysh theory.
Although the Keldysh theory reliably reproduces the main features of the TDDFT calculation, we find some deviations between results
by the two theories. The origin of the differences is examined in detail.
\PACS{
      {PACS-key}{discribing text of that key}   \and
      {PACS-key}{discribing text of that key}
     } 
} 
\maketitle
\section{Introduction}
\label{intro}

Nonlinear optical phenomena induced by high-intensity ultrashort laser pulses have been actively
investigated in wide and diverse fields in basic sciences, technologies, industries, and medicines.
Interests on and capabilities of the nonlinear optical phenomena are still rapidly growing owing to the
further developments of the laser technologies.
In the present work, we focus on interactions of femtosecond laser pulses with dielectrics at an intensity
range around the damage threshold. Above the threshold, the interaction is directly related to the
understanding of nonthermal laser processing that has been actively explored in last two decades\cite{lorazo2003,balling2013,kumada2014}.
Below the threshold, an ultrafast control of electron dynamics in a medium, even within a cycle of 
the laser pulse, has been actively investigated in recent years\cite{shugaev2016}. Utilizing attosecond metrology, 
it has become possible to measure electron dynamics in attosecond time-scale\cite{Krausz2009,calegari2016}. 
In spite of such rapid, mostly experimental progresses, theoretical description and understanding of the process 
is still under development because of highly nonlinear and complex nature of the interaction between the strong 
pulsed electric field of the laser pulse and electrons in dielectrics.

A fundamental process that takes place first in the interaction between a high-intensity ultrashort laser pulse and dielectrics
is an excitation of electrons from valence bands to conduction bands. 
While one photon absorption is the major process for a laser pulse of frequency above the optical bandgap,
two nonlinear processes, multiphoton absorption and tunneling ionization, are important for a pulse of frequency below the bandgap.
As a theory that is capable of describing these mechanisms, the Keldysh theory has been developed\cite{keldysh1965} and 
widely utilized\cite{balling2013}. Although it has been quite successful to explain various features of the nonlinear electronic
excitations in atoms, molecules, and solids, it includes approximations in the derivation and empirical
parameters in the formula. Developments of theories and calculations with less approximations and/or 
less empirical parameters are desirable.

As a theory and computational method that meets the above request, time-dependent density functional theory
(TDDFT)\cite{Runge1984,Ullrich2012,Marques2012}, an extension of ordinary density functional theory so as to describe electronic excitations and dynamics,
provides a reasonable starting point. It has been well recognized that TDDFT combined with linear response
theory has been quite successful to describe electronic excitations of molecules in the first-principles level\cite{Casida2009,Laurent2013}. 
Solving the time-dependent Kohn-Shan (TDKS) equation in real time, it is capable of describing highly nonlinear 
interactions between high-intensity ultrashort laser pulses and electrons in various matters without any empirical parameters
as long as the functional adopted in the calculation does not include parameters for a specific purpose.

Our group has been developing a computational method to describe electron dynamics based on TDDFT,
solving the TDKS equation in real time and real space\cite{Yabana1996}. The program is developed as an open source
software, SALMON (Scalable Ab-initio Light-Matter simulator for Optics and Nanoscience)\cite{SALMON_paper2018}, and can be
downloadable from our website \cite{SALMON_web}. For electron dynamics in crystalline solids,
it has been applied to calculations of dielectric functions\cite{Bertsch2000}, coherent phonon generations\cite{Shinohara2010,Shinohara2012}, 
dielectric breakdown\cite{Otobe2008}, and so on\cite{Yabana1999,Otobe2009-2,Sato2014PRB,Wachter2014,Schultze2014}.
Usefulness of electron dynamics calculations based on TDDFT is further enhanced by combining them with Maxwell
equations that describe the propagation of light wave. We have developed the multiscale simulation method \cite{Yabana2012}
and have been successfully applied for the analyses of attosecond experiments in crystalline solids \cite{Sommer2016,Lucchini2016}
and also for the initial stage of nonthermal laser processing of transparent dielectrics\cite{Lee2014,Sato2015}. 

In this paper, we report a systematic investigation of the energy transfer from a pulsed electric field to electrons
in a unit cell of crystalline solids, solving the TDKS equation in real time. Frequency as well as intensity of the 
pulsed electric field is varied to investigate the mechanism. We compare the calculated results with the Keldysh theory. 
Calculations are carried out for three simple materials, silicon, diamond, and $\alpha$-quartz (SiO$_2$). 
Silicon and diamond are covalent materials with different optical bandgap. $\alpha$-quarts is an ionic material with 
a wide bandgap. There are several purposes of the present analyses.
We expect the results of systematic calculations and their comparison with the formula of the Keldysh theory
will be useful to understand the energy transfer mechanisms and to clarify the usefulness and the limitation of the
formula of the Keldysh theory. We also expect to find a novel nonlinear interaction that are ignored in the Keldysh theory.

The organization of the paper is as follows: In Section 2, we provide an overview of the formalism and the computational method.
In Section 3, calculated results are presented and are compared with the Keldysh theory. From the calculated results,
excitation mechanisms are discussed in detail. In Section 4, concluding remarks are presented.

\section{Theoretical Method}
\label{sec:1}

In this section, our theoretical and computational method based on TDDFT is briefly explained. 
Full explanation has been given in our previous publications\cite{SALMON_paper2018,Otobe2008,Yabana2014}.
Practical calculations are carried out using SALMON \cite{SALMON_web}, an open source
software developed in our group.

To describe interaction between a high-intensity pulsed light and electrons in dielectrics,
we consider electron dynamics induced by the pulsed electric field in a unit cell of a crystalline solid. 
Since the wavelength of the light is much longer than a typical spatial scale of electron motion, 
we assume a dipole approximation treating the electric field of the pulsed light as a spatially uniform field.
We use a velocity gauge for the description, using a vector potential $\vec A(t)$ to express 
the pulsed electric field $\vec E(t)$,
\begin{equation}
\vec E(t) = - \frac{1}{c} \frac{d}{dt} \vec A(t).
\label{eq:EA}
\end{equation}
The Kohn-Sham Hamiltonian in this gauge has the spatial lattice periodicity at each time.
Therefore, we may apply the Bloch theorem at each time and describe the electron dynamics
using Bloch orbitals, $u_{b \vec k}(\vec r, t)$, where $b$ is the band index and $\vec k$ is the
crystalline momentum. The TDKS equation for the Bloch orbital is given by
\begin{eqnarray}
  i\hbar\frac{\partial}{\partial t}u_{b \vec k}(\vec r,t) &=&
  \Biggl[
    \frac{1}{2m_e}\left\{-i\hbar\nabla+\hbar\vec k + \frac{e}{c}\vec A(t) \right\}^2 + v_{H}(\vec r,t)
    \Biggr. \nonumber \\
    && \Biggl.  + v_{xc}(\vec r,t) + \hat v_{ion}
  \Biggr] u_{b \vec k}(\vec r,t),  \label{KS-u}
\end{eqnarray}
where the Kohn-Sham Hamiltonian consists of four terms. The kinetic energy operator with the
electron mass $m_e$ includes the vector potential $\vec A(t)$.
Remaining three potentials, Hartree potential $v_{H}(\vec r,t)$, exchange-correlation potential $v_{xc}(\vec r,t)$, 
and ionic potential $\hat v_{ion}$ have the lattice periodicity at each time.
For the ionic potential, we utilize norm-conserving pseudopotential \cite{Troullier1991} with separable 
approximation \cite{Kleinman1982}. 

For a reliable description of laser-solid interaction, it is important to utilize energy functional
that reproduces energetic features of each material quantitatively.
We employ the TB-mBJ potential\cite{Becke2006,Tran2009}, which is a meta-GGA potential that depends on density, gradient of 
the density, and kinetic energy density, and is known to reproduce bandgap energies of dielectrics
systematically\cite{Tran2009}. The TB-mBJ potential includes one parameter $c$ for which there is a 
recommended procedure to determine the value \cite{Tran2009}. We instead determine it to accurately
reproduce the energy of the indirect bandgap of materials. The tuned values are 1.04 for the silicon, 
1.22 for the diamond and 1.00 for the $\alpha$-quartz, where the calculated indirect band gap energies 
are 1.26, 5.48 and 7.6 eV, respectively.

We use the following time profile of cosine-squared envelope shape for the vector potential, which gives the
electric field of the applied laser pulse through Eq. (\ref{eq:EA}),
\begin{eqnarray}
  \vec A(t) = -\frac{A_0}{\omega} \cos^2 \left(\frac{\pi t}{T} \right) \cos(\omega t) \vec e_c  \label{At}, 
  \hspace{5mm} ( 0 < t < T),
\label{eq:At}
\end{eqnarray}
where $\omega$ is the average frequency and $T$ is the duration of the pulsed electric field.
The polarization vector is chosen to be $c$ direction for all systems.
Due to the structural symmetry, we expect the results do not depend strongly on the polarization 
direction for silicon and diamond. Results for $\alpha$-quartz may differ for different directions, 
but we consider that we can at least investigate qualitative features. 

We solve Eq. (\ref{KS-u}) in time domain. Using calculated Bloch orbitals $u_{b\vec k}(\vec r,t)$,
we can calculate the macroscopic electric current density $\vec J(t)$ that is the spatial average of the microscopic
electric current density $\vec j(\vec r,t)$ averaged over the unit cell volume. It is given by
\begin{eqnarray}
  \vec j (\vec r, t) &=& \frac{-e}{2m_e}\sum_{b,\vec k}
  \Biggl[ u^*_{b \vec k}(\vec r,t)
  \left\{-i\hbar\nabla+\hbar\vec k + \frac{e}{c}\vec A(t) \right\}
  u_{b \vec k}(\vec r,t) \Biggr.  \nonumber \\
  \Biggl. &&
  + c.c.   \Biggr] + \vec j_{PS},
\end{eqnarray}
where $\vec j_{PS}$ is a current contribution coming from the nonlocal pseudpotential.

\begin{figure*}[h]
\centering
\resizebox{0.8\textwidth}{!}{\includegraphics{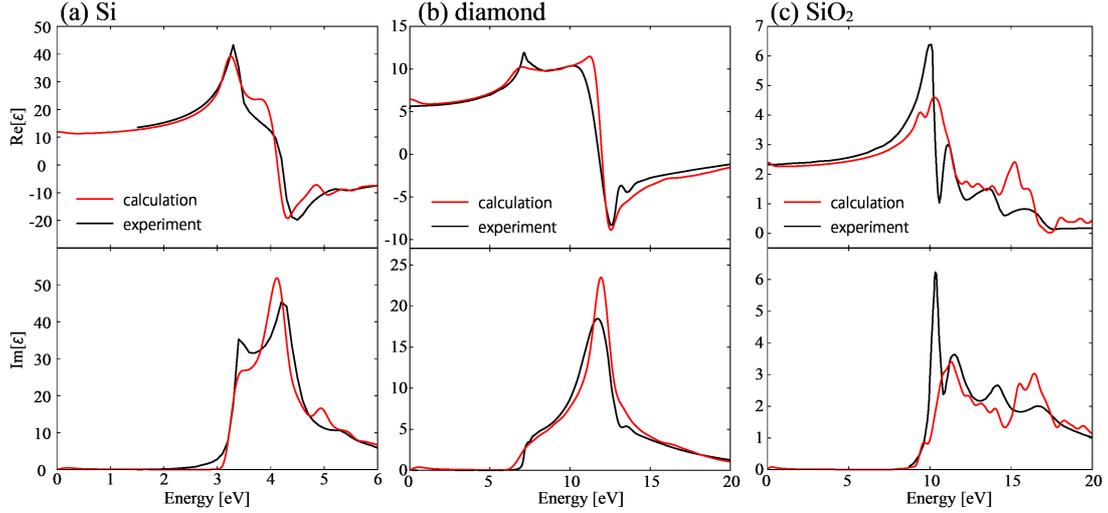}}
\caption{
  (Color online) Calculated real (upper panel) and imaginary (lower panel) parts of dielectric functions for (a) silicon, (b) diamond and (c) $\alpha$-quartz (red lines). 
  The experimental measurements\cite{aspnes1983,edwards1985handbook,philipp1966} are plotted in the black lines for comparison.}
\label{fig-eps}
\end{figure*}

A principal quantity that we will investigate in the next section is the electronic excitation
energy $\Delta E_{\rm ex}(t)$ at time $t$. This should be equal to the energy transfer from 
the applied pulsed electric field to electrons in a unit cell of crystalline solid. We evaluate it from 
the work that the pulsed electric field $\vec E(t)$ do for the electrons in the unit cell,
\begin{equation}
  \Delta E_{\rm ex}(t) = \int^t \vec E(t') \cdot \vec J(t') dt'.
\label{eq:Eex}
\end{equation}
If the exchange-correlation potential is derived from an energy density functional, we can show
that the work done by the pulsed electric field is equal to the difference between the total electronic energy 
at time $t$ and the ground state energy, as we showed in Ref. \cite{Sato2015JCP}.
However, since the TB-mBJ exchange-correlation potential that we use in our calculation is not derived from
any energy density functional, we calculate $\Delta E_{\rm ex}(t)$ using Eq. (\ref{eq:Eex}).

In practical calculations, we use a uniform spatial grid in the Cartesian coordinate system to express 
the Bloch orbitals and the potential terms.
The spacial grid sizes $\Delta x$, $\Delta y$, $\Delta z$ as well as numbers of $k$-points and time step 
size d$t$ are determined so that the calculated results converge.
The determined parameters of the grid and step sizes are
$\Delta x$=$\Delta y$=$\Delta z$=0.34\AA, $12^3$ of $k$-points and d$t=$0.48 attosecond(as) for the silicon, 
$\Delta x$=$\Delta y$=$\Delta z$=0.22\AA, $12^3$ of $k$-points and d$t=$0.48 as for the diamond ,
and $\Delta x$=0.18\AA, $\Delta y$=0.30\AA, $\Delta z$=0.19\AA, $4^3$ of $k$-points and d$t=$0.24 as for the $\alpha$-quartz.

To confirm the reliability of the exchange-correlation potential employed in our calculation, 
we calculated real and imaginary parts of the dielectric function of the three systems. 
They are shown in Fig.\ref{fig-eps}.
In the calculation, we use the same numerical scheme solving the TDKS equation
in real time \cite{Bertsch2000,Yabana2012}, as described above.
The overall shapes of the dielectric functions are in reasonable agreement with the measurements. 
Especially, the rising positions of the first main peaks in the imaginary part, which are around 
3.1eV (silicon), 6.5 eV (diamond) and 9.0 eV ($\alpha$-quartz), are well reproduced.
In our calculations, atomic positions are frozen at their equilibrium positions  during the calculations
so that our calculations do not include any excitations below the direct bandgap energy assisted by phonons.

\section{Results and Discussion}

\begin{figure}[h]
\centering
\resizebox{0.95\linewidth}{!}{\includegraphics{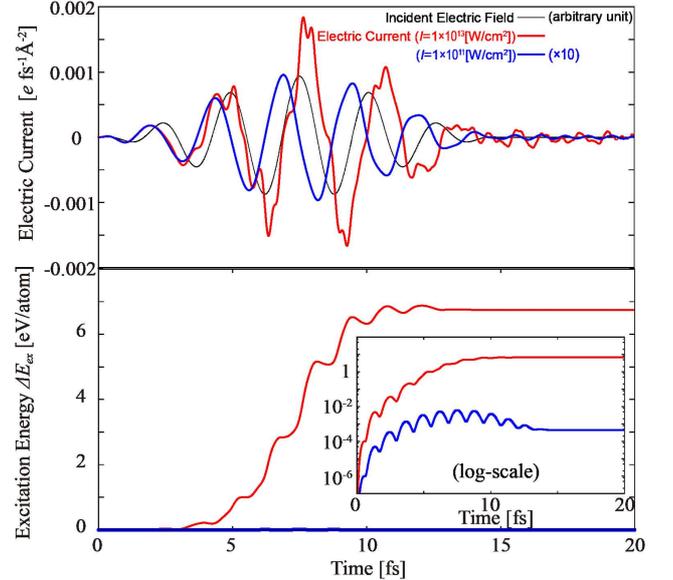}}
\caption{
  (Color online) A typical time evolution calculation in silicon. In upper panel, applied pulsed electric field is shown by black-solid line.
  Induced electric current density is shown at two intensities, blue line for $I=10^{11}$ W/cm$^2$ (multiplied by a factor 10) and
  red line for $I=10^{13}$ W/cm$^2$. In lower panel, the electronic excitation energy in a unit cell is shown.
  }
\label{fig-time}
\end{figure}

In the following, we show calculated results of electronic excitations in a unit cell of crystalline solids induced by a pulsed laser light 
whose electric field is given by Eqs. (\ref{eq:EA}) and (\ref{eq:At}).
The duration of the pulse is set to $T=$15 fs (5.5 fs at the FWHM of the envelope), while the amplitude and the frequency are varied. 
To indicate the intensity of the pulsed laser light, we use the intensity $I$ defined by $I=c\epsilon_0 E_{\rm max}^2/2$ where 
$E_{\rm max}$ is the maximum amplitude of the pulsed electric field.
We note that this is the relation connecting the amplitude of the electric field and the intensity of the pulsed laser light
in the vacuum, not in the medium.

In Fig. \ref{fig-time}, we show a typical electron dynamics calculation in a crystalline silicon. 
In the upper panel, the time profile of the pulsed electric field $E(t)$ (black line) and 
the induced electric current density $J(t)$ is shown for two cases of different intensities, $I=10^{11}$ W/cm$^2$ and $I=10^{13}$ W/cm$^2$. 
Note that the current density at $I=10^{11}$ W/cm$^2$ is multiplied by a factor of 10.
The frequency of the pulsed electric field is set to $\hbar\omega=1.55$ eV.
At the intensity of $I=10^{11}$W/cm$^2$, the electric current density shows a similar time profile to the pulsed electric field,
except for a phase difference of $\pi/2$. 
This indicates a linear dielectric response of silicon for a weak electric field at the frequency below the bandgap energy.
At the intensity of $I=10^{13}$W/cm$^2$, the time profile of the electric current density is strongly distorted and is very different 
from the lower intensity case. This indicates a strongly nonlinear response of electrons to the electric field.

In the lower panel, the electronic excitation energy $\Delta E_{\rm ex}(t)$ defined by Eq. (\ref{eq:Eex}) is shown. 
At $I=10^{11}$ W/cm$^2$, the excitation energy is substantially small. 
In the log-scale shown in the inset, the excitation energy is visible during the irradiation of the pulsed electric field
and becomes very small after the pulsed electric field ends, since the electronic state goes back almost to the ground state.
Contrarily, substantial excitation energy, about 6 eV/atom, is seen at the intensity of $I=10^{13}$ W/cm$^2$
due to highly nonlinear interactions. 

\begin{figure*}[h]
\centering
\resizebox{0.9\textwidth}{!}{\includegraphics{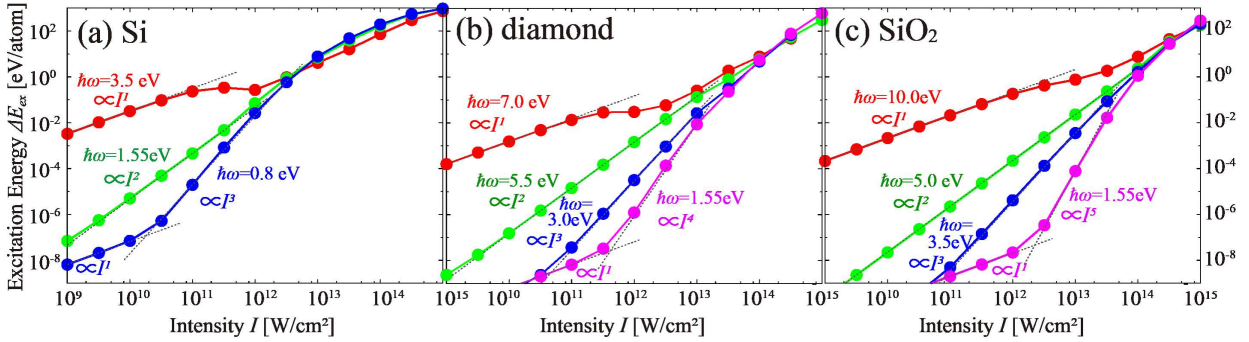}}
\caption{
  (Color online) Electronic excitation energies in TDDFT calculation as a function of the intensity of the applied laser pulse with several frequencies 
  (a) for silicon, (b) for diamond, and (c) for $\alpha$-quartz.}
\label{fig-EIw-simu}
\end{figure*}

In Fig.\ref{fig-EIw-simu}, we summarize our calculated results of the electronic excitation energy, $\Delta E_{\rm ex}(t)$, 
at a time after the field ends.
Panels (a), (b), and (c) show results of silicon, diamond, and $\alpha$-quartz, respectively. 
In each panel, the excitation energy is shown as a function of the intensity of the applied laser pulse $I$, 
for several different frequencies $\omega$. 
From the figure, we find the following characteristic features.

\begin{figure*}[h]
\centering
\resizebox{0.9\textwidth}{!}{\includegraphics{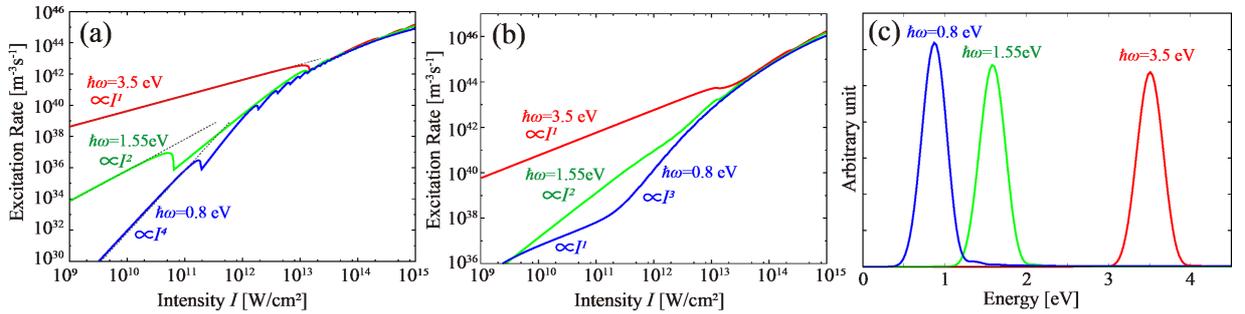}}
\caption{(Color online) Excitation rates of silicon by the Keldysh theory as a function of the intensity of the applied laser pulse with several chosen photon energies,
  (a) for a continuous wave, (b) convoluted using power spectra of pulsed electric field. Power spectra that is used as weight functions are shown in (c).
  }
\label{fig-keldysh}
\end{figure*}

\begin{description}

\item[(1)]
There is an intersection point in each plot at which the excitation energies of different frequencies merge into 
the same value at a certain intensity.
The intensity of the intersection point depends on the material:
$3.2 \times 10^{12}$ W/cm$^2$ in silicon, $1.0 \times 10^{14}$ W/cm$^2$ for diamond, and $1.0 \times 10^{15}$ W/cm$^2$ for $\alpha$-quartz. 

\item[(2)]
At intensities below the intersection point, the excitation energies show perturbative behavior and
well fit by $I^N$ where $N$ is an integer. The integer number $N$ coincides with the number of 
photons required to exceed the bandgap energy. There are, however, exceptions as described in (3) below.

\item[(3)]
The integer number $N$ is sometimes lower than the number of photons required to across the direct bandgap. 
In silicon at $\hbar\omega = 0.8$ eV, four photons are required to across the direct bandgap, while the excitation
energy is consistent with $N=3$. At very low intensity region where the excitation energy is below 
$10^{-6}$ eV/atom and using low frequency electric field, the energy transfer shows $N=1$ behavior in all three materials.

\item[(4)]
At intensities above the intersection point, the excitation energy does not depend much on the frequency, as seen
in the case of silicon. The slope becomes small compared with the low intensity region.

\item[(5)]
Using electric fields of frequencies above the bandgap energy, excitation energy curves show somewhat a planar or 
a small gradient region just below the intersection point and are not fit well by $I^1$ curve in all three materials.

\end{description}

Among these behaviors, items (1), (2), and (4) are consistent with descriptions using the so-called Keldysh theory of ionization in solids \cite{keldysh1965}. 
The Keldysh theory describes strong field ionization phenomena with full inclusion of nonlinearity for an irradiation
of a continuum wave field and has been widely used for analyses of experiments.
To make a quantitative comparison of our TDDFT results with the Keldysh theory, we show the excitation rate of silicon 
by the Keldysh theory in Fig.\ref{fig-keldysh}(a) where the direct bandgap energy value of silicon is set to 3.09 eV, 
and effective masses of electrons and holes to 1.06 $m_e$ and 0.59 $m_e$, respectively\cite{green1990}.
The figure shows that the Keldysh theory reproduces the overall trend of Fig.\ref{fig-EIw-simu} very well, confirming that 
fundamental essence of the nonlinear interaction is certainly captured in the Keldysh theory.
The intersection point is related to the change from a perturbative to a tunneling regime in which the border of two regimes are
divided by the so-called Keldysh parameter, $\gamma$.  At low intensity region where the Keldysh parameter is much 
lower than unity, multiphoton absorption is expected. At high intensity region where $\gamma$ is larger than unity, we expect tunneling and
above-barrier ionizations dominate and the frequency of the field becomes less important.

However, the result of the Keldysh theory does not reproduce the features of items (3) and (5).
To understand the reason of the item (3), we make a smoothing of the ionization rate of the Keldysh theory utilizing the frequency 
distribution of the pulsed electric field employed in the TDDFT calculation.
We show in Fig.\ref{fig-keldysh}(b) the excitation rate that is obtained by convoluting the excitation rate of Fig.\ref{fig-keldysh}(a) 
with the frequency distribution of the pulsed electric field that is shown in Fig.\ref{fig-keldysh}(c).
As seen from Fig.\ref{fig-keldysh}(b), the slope of $\hbar\omega$=0.8 eV now becomes consistent with $I^3$ in the intensity
region from $5 \times 10^{11}$ W/cm$^2$ to $1.0 \times 10^{13}$ W/cm$^2$.
We also observe an appearance of the slope of $I^1$ below $5 \times 10^{11}$ W/cm$^2$. 
Paying attention to details of Fig.\ref{fig-keldysh}(a), small humps appear in the Keldysh theory calculations below the intersection point 
due to the ponderomotive effect, the increased bandgap energy by the oscillating field. In the TDDFT calculation, such humps are not observed.
The humps disappeared by the frequency averaging, indicating that the effect cannot be observed if one employs such short pulses as used here. 

\begin{figure*}[h]
\centering
\resizebox{0.7\textwidth}{!}{\includegraphics{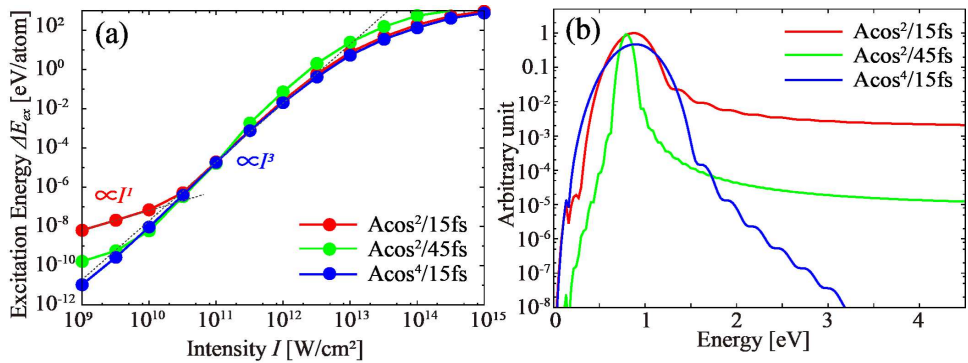}}
\caption{(Color online) (a) Excitation energies for silicon using pulsed electric fields of three different envelopes and a common frequency, $\hbar\omega$=0.8 eV.
(b) Power spectra of the fields are shown. Notation ``Acos$^2$/15fs'', for example, indicates the use of the envelope function with the square of cosine with a duration of 15 fs.
 \label{fig-EI-compare-pulse}}
\end{figure*}

To further clarify the origin of the $I^1$ slope, we show  in Fig.\ref{fig-EI-compare-pulse}(a) the excitation energies in the TDDFT calculation 
using different pulse shapes. In the figure, three calculations are shown: Results using the pulse shape of Eq. (\ref{eq:At}) with 
$T=15$ fs by red curve, which corresponds to the result shown in Fig. \ref{fig-EIw-simu}, results using the pulse shape of Eq. (\ref{eq:At}) 
with $T=45$ fs by green curve, and the result using the pulse shape with cosine to the fourth power envelope with $T=15$ fs by blue curve.
Figure \ref{fig-EI-compare-pulse}(b) show spectral distributions of the pulses. 
As seen in the figure, the field using the pulse shape of Eq. (\ref{eq:At}) includes a small but long tail towards the high frequency.
At very low intensity region where the excitation energy is very small, one-photon absorption process that can take place by the high frequency
component of the pulse is dominated. This explains the appearance of $I^1$ behavior at low excitation region in all three materials.
Such behavior is not seen if we use a pulse of $\cos^4$ envelope, since it does not have the high frequency component.
In summary, we need to be careful in choosing the shape of the pulse envelope when we are interested in nonlinear electronic excitations
in wide bandgap materials and the electronic excitation energy is very small.


\begin{figure*}[h]
\centering
\resizebox{0.9\textwidth}{!}{\includegraphics{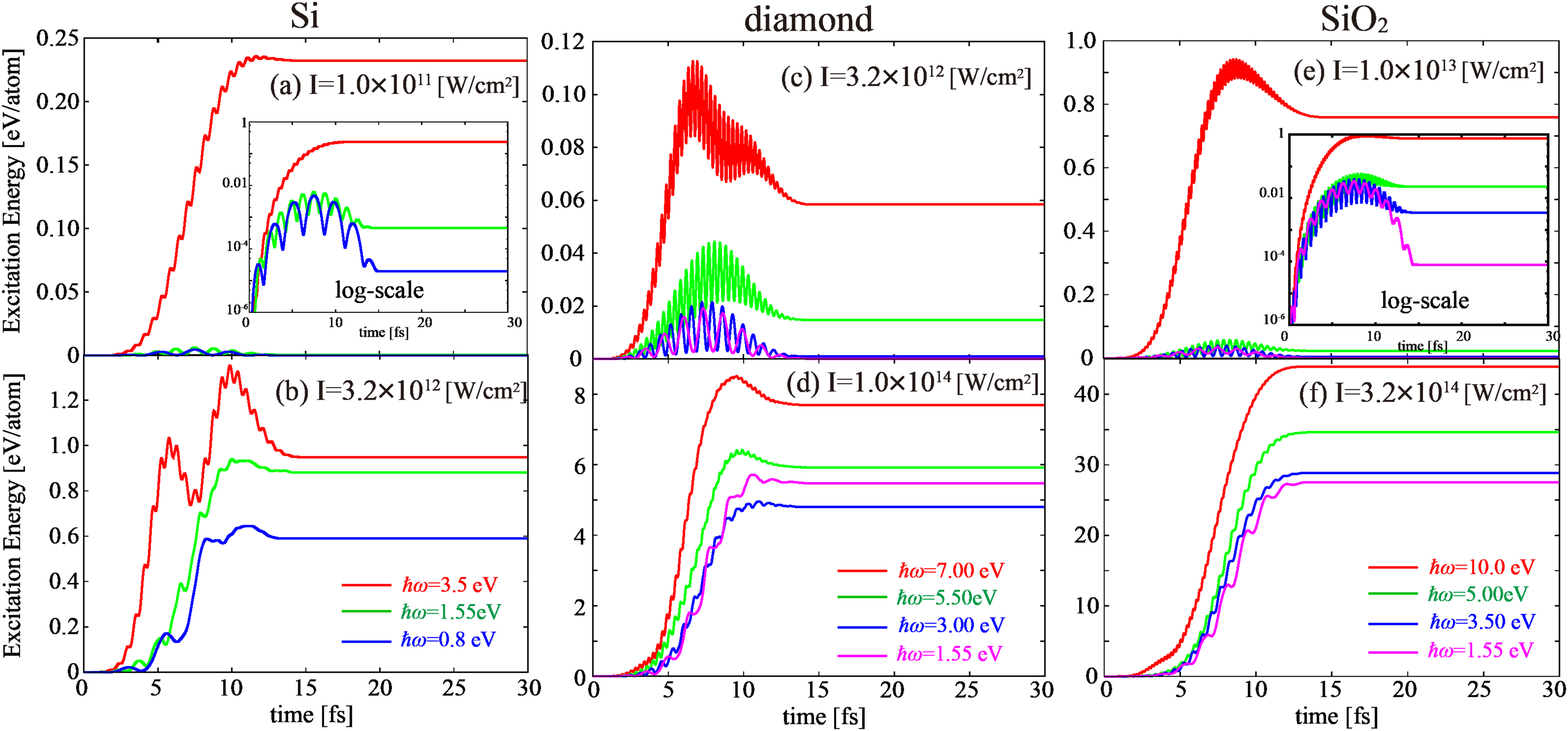}}
\caption{(Color online) Temporal development of excitation energies at several frequencies for silicon ((a)and (b)), diamond ((c)and (d)), and $\alpha$-quartz ((e)and (f)).
  Results of low field intensity below the intersection region in Fig.\ref{fig-EIw-simu} is shown in the upper panels,
  and those of the intensity around the intersection are shown in the lower panels.
}
\label{fig-EtI-simu}
\end{figure*}

\begin{figure}[h]
\centering
\resizebox{0.4\textwidth}{!}{\includegraphics{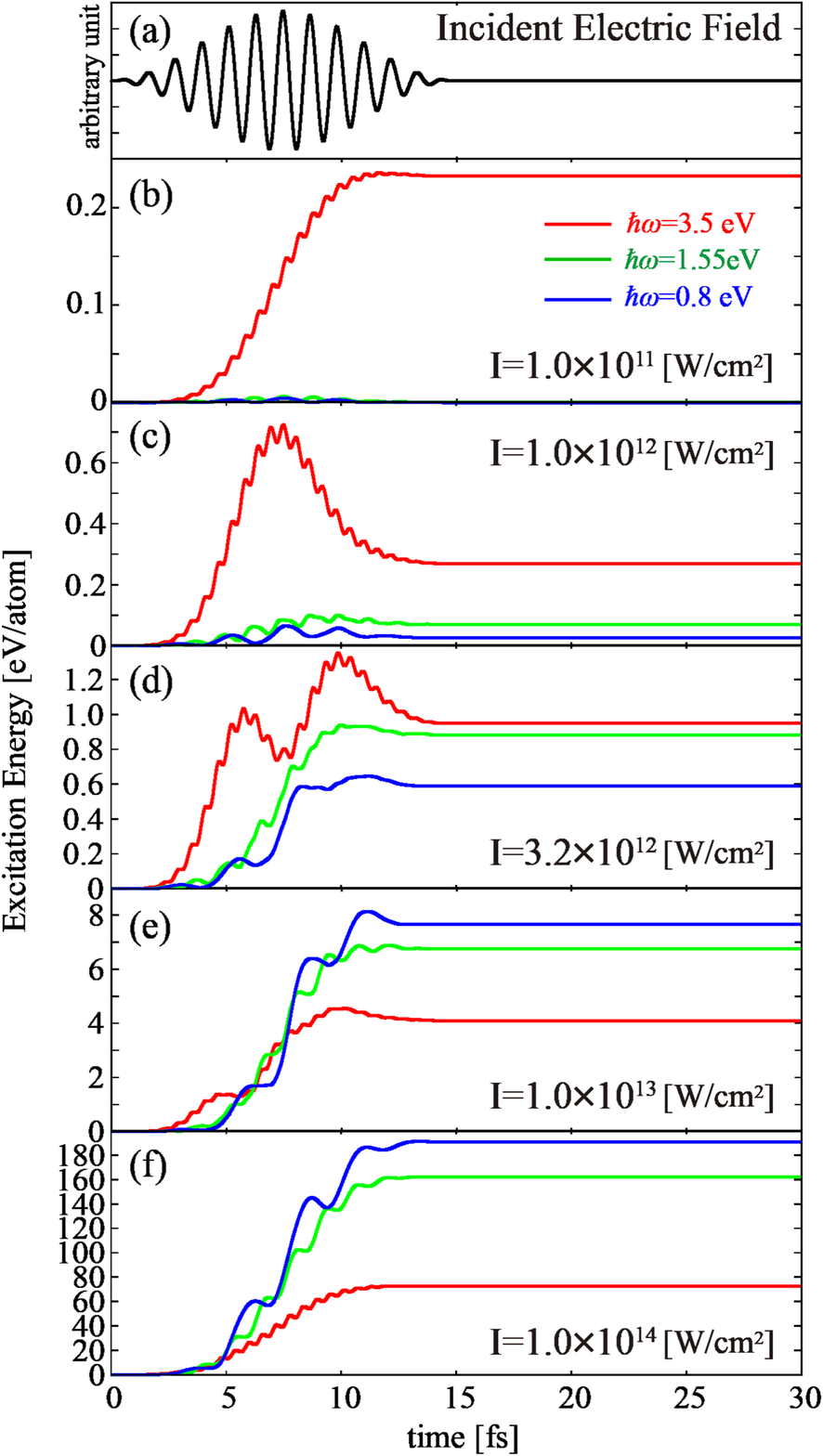}}
\caption{(Color online) Temporal development of excitation energies of silicon during and after the pulse irradiation using the field intensities below and above 
the intersecting region in Fig.\ref{fig-EIw-simu}. }
\label{fig-EtI-simu-Si}
\end{figure}

We finally discuss the item (5), an appearance of a planar or a small gradient region
in the excitation energy curve just below the intersection point at frequencies above the bandgap energy,
which are seen in all three materials. In the Keldysh theory calculation, a small dip is seen and originates
from the effect of the ponderomotive energy. However, the magnitude of the dip is much smaller than that in
the TDDFT calculation.

In order to figure out the mechanisms, we look at in detail the temporal development of the electronic excitations, 
i.e. $\Delta E_{\rm ex}(t)$ as a function of time.
Fig.\ref{fig-EtI-simu} shows them at intensities below and around the intersection region of each material. 
Excitation energies at the intensity below (above) the intersection region is shown in the upper (lower) panels, respectively.

As a general trend, the temporal development of the excitation energy consists of two kinds of profiles:
One is the excitation profile that behaves like $E(t)^2$ and is seen during the irradiation of the pulsed electric field.
It is dominated when the frequency is below the bandgap and the intensity is not very strong. 
This is a virtual energy transfer from the laser pulse to electrons. The energy is returned back to the laser pulse and 
disappears after the pulsed field ends.
The other is a resonant excitation that shows gradual increase of excitation energies during the irradiation of the pulsed 
electric field and persists after the pulsed field ends. This process takes place when the frequency is above the bandgap
energy or when the intensity of the field is very strong so that nonlinear excitations such as multiphoton absorption or
tunneling ionization take place.

There appears an intriguing oscillation in the electronic excitation in silicon at around the intersection region,
which is shown in Fig.\ref{fig-EtI-simu}(b) with $I=3.2\times 10^{12}$ W/cm$^2$ and $\hbar\omega = 3.5$ eV.
The period of the oscillation is much larger than the period of the field itself.
We consider that this behavior can be attributed to the Rabi oscillation under a strong field \cite{michael2016}.
The period of the oscillation is consistent with the crude estimate $\hbar \Omega = 0.9$ eV,
which corresponds to period of 4.6 fs, using a formula 
$\hbar\Omega = 2 \mu E$  with the typical dipole matrix element $\mu = $1.8 $e${\AA} from our calculation and 
the electric field amplitude $E=0.25$ V/{\AA} that is taken as a half of the maximum amplitude $E_{\rm max}$ 
at $I=3.2 \times 10^{12}$ W/cm$^2$.
The period at $I=1.0 \times 10^{12}$ W/cm$^2$ estimated in the same way is 8.5 fs, that is also consistent
with Fig.\ref{fig-EtI-simu-Si}(c) as will be seen later. 


The appearance of the Rabi oscillation indicates that substantial changes occur in occupation number
distributions in the valence and conduction orbitals during the irradiation of the laser pulse. 
Such occupation changes are expected to produce strong nonlinear effects even when the laser frequency is 
above the bandgap energy and the one-photon absorption process dominates in the excitation.
We consider the item (5), appearance of a planar or a small gradient region, is related to the nonlinearity 
caused by the occupation change.
Since substantial excitations take place at the beginning of the irradiation of the pulsed electric field, 
the electronic excitation saturates during the irradiation and the energy transfer from the pulsed electric 
field to electrons becomes smaller than a simple estimate from the one-photon absorption.

To examine the excitation mechanisms in more detail, we show in Fig. \ref{fig-EtI-simu-Si} 
the time evolution of electronic excitations at several intensities. 
Looking at curves of $\hbar\omega = 3.5$ eV, a simple absorption process is seen at low intensity, 
$I=1.0\times 10^{11}$ W/cm$^2$, as shown in panel (a).
Increasing the intensity toward the intersection region, $I=1.0\times 10^{12}$ (in panel (b)) and 
$3.2\times 10^{12}$ (in panel (c)) W/cm$^2$, one- and two-cycle oscillation are observed. 
These periods correspond roughly to the period of Rabi oscillation as mentioned before.
Further increasing the intensity above $10^{13}$ W/cm$^2$ (in panels (d) and (e)), the oscillation no more appears.
At these intensities, the excitation energy after the pulse ends is smaller than the cases of lower frequencies,
which was also observed in Fig. \ref{fig-EIw-simu}(a).
We consider this suppression of the electronic excitation is also related to the saturation effect of the absorption. 

\section{Concluding Remarks}

We have presented a systematic investigation for electronic excitations in dielectrics under an irradiation of
an intense and ultrashort laser pulse employing first-principles time-dependent density functional theory.
Freqnecy and intensity dependences are examined in detail for three typical materials with 
different bandgap energies and cohesion mechanisms, silicon, diamond, and $\alpha$-quartz. 
The calculated results are compared with the excitation rates by the Keldysh theory. 
It is found that basic features of the excitation is well described by the Keldysh theory. 
However, we find there are several features that require careful considerations.
The frequency distribution of the pulsed electric field, in particular the high frequency component, should be 
carefully examined since it may change the number of photons required to across the bandgap 
when the excitation rate is very low.
We also find an appearance of complex nonlinear behavior for strong lasers pulse with the frequency above 
the bandgap. The Rabi-like oscillation appears in the temporal development of the excitation energy,
and a suppression in the electronic excitation is found. We attribute them to the nonlinearity originated from 
the saturation effect in the occupation.


%




\section*{Acknowledgement}

This research was supported by JST-CREST under grant number JP-MJCR16N5, and by
MEXT as a priority issue theme 7 to be tackled by using Post-K Computer, and by JSPS KAKENHI
Grant Numbers 15H03674.
Calculations are carried out at Oakforest-PACS at JCAHPC under the support by Multidisciplinary
Cooperative Research Program in CCS, University of Tsukuba.

\section*{Authors contributions}
AY performed the simulations. Both authors (AY and KY) contributed to the manuscript.

%
%
%


\bibliographystyle{unsrt}
\bibliography{bibf/yabana,bibf/theory,bibf/laser,bibf/tddft,bibf/other_misc_1}

\end{document}